\input harvmac
\input epsf.tex
\def\ev#1{\langle#1\rangle}

\noblackbox
\def\IL{\relax{\rm I\kern-.18em L}}
\def\IH{\relax{\rm I\kern-.18em H}}
\def\IR{\relax{\rm I\kern-.18em R}}
\def\IC{\relax\hbox{$\inbar\kern-.3em{\rm C}$}}
\def\IZ{\relax\ifmmode\mathchoice
{\hbox{\cmss Z\kern-.4em Z}}{\hbox{\cmss Z\kern-.4em Z}}
{\lower.9pt\hbox{\cmsss Z\kern-.4em Z}} {\lower1.2pt\hbox{\cmsss
Z\kern-.4em Z}}\else{\cmss Z\kern-.4em Z}\fi}
\def\CM {{\cal M}}

\def\CP {{\cal P }}

\def\CO {{\cal O}}


\def\det{{\rm det}}
\def\Tr{{\rm Tr}}

\font\manual=manfnt \def\dbend{\lower3.5pt\hbox{\manual\char127}}

\def\IZ{\relax\ifmmode\mathchoice
{\hbox{\cmss Z\kern-.4em Z}}{\hbox{\cmss Z\kern-.4em Z}}
{\lower.9pt\hbox{\cmsss Z\kern-.4em Z}} {\lower1.2pt\hbox{\cmsss
Z\kern-.4em Z}}\else{\cmss Z\kern-.4em Z}\fi}
\def\half {{1\over 2}}

\lref\BarnesJJ{
  E.~Barnes, K.~A.~Intriligator, B.~Wecht and J.~Wright,
  ``Evidence for the strongest version of the 4d a-theorem, via a-maximization along RG flows,''
Nucl.\ Phys.\ B {\bf 702}, 131 (2004).
[hep-th/0408156].
}
\lref\IW{
  K.~A.~Intriligator and B.~Wecht,
  ``The Exact superconformal R symmetry maximizes a,''
Nucl.\ Phys.\ B {\bf 667}, 183 (2003).
[hep-th/0304128].
}
\lref\GreenDA{
  D.~Green, Z.~Komargodski, N.~Seiberg, Y.~Tachikawa and B.~Wecht,
  ``Exactly Marginal Deformations and Global Symmetries,''
JHEP {\bf 1006}, 106 (2010).
[arXiv:1005.3546 [hep-th]].
}
\lref\KSLM{
  D.~Kutasov and A.~Schwimmer,
  ``Lagrange multipliers and couplings in supersymmetric field theory,''
Nucl.\ Phys.\ B {\bf 702}, 369 (2004).
[hep-th/0409029].
}
\lref\DimoftePD{
  T.~Dimofte and D.~Gaiotto,
  ``An E7 Surprise,''
[arXiv:1209.1404 [hep-th]].
}
\lref\FreedmanRD{
  D.~Z.~Freedman and H.~Osborn,
  ``Constructing a c function for SUSY gauge theories,''
Phys.\ Lett.\ B {\bf 432}, 353 (1998).
[hep-th/9804101].
}
\lref\CohenSQ{
  A.~G.~Cohen and H.~Georgi,
  ``Walking Beyond The Rainbow,''
Nucl.\ Phys.\ B {\bf 314}, 7 (1989).
}
\lref\LeighEP{
  R.~G.~Leigh and M.~J.~Strassler,
  ``Exactly marginal operators and duality in four-dimensional N=1 supersymmetric gauge theory,''
Nucl.\ Phys.\ B {\bf 447}, 95 (1995).
[hep-th/9503121].
}
\lref\ZamolodchikovGT{
  A.~B.~Zamolodchikov,
  ``Irreversibility of the Flux of the Renormalization Group in a 2D Field Theory,''
JETP Lett.\  {\bf 43}, 730 (1986), [Pisma Zh.\ Eksp.\ Teor.\ Fiz.\  {\bf 43}, 565 (1986)].
}
\lref\BuicanTY{
  M.~Buican,
  ``A Conjectured Bound on Accidental Symmetries,''
Phys.\ Rev.\ D {\bf 85}, 025020 (2012).
[arXiv:1109.3279 [hep-th]].
}
\lref\ShifmanZI{
  M.~A.~Shifman and A.~I.~Vainshtein,
  ``Solution of the Anomaly Puzzle in SUSY Gauge Theories and the Wilson Operator Expansion,''
Nucl.\ Phys.\ B {\bf 277}, 456 (1986), [Sov.\ Phys.\ JETP {\bf 64}, 428 (1986)], [Zh.\ Eksp.\ Teor.\ Fiz.\  {\bf 91}, 723 (1986)]..
}
\lref\Kutasov{
  D.~Kutasov,
  ``A Comment on duality in N=1 supersymmetric nonAbelian gauge theories,''
Phys.\ Lett.\ B {\bf 351}, 230 (1995).
[hep-th/9503086].
}
\lref\ErkalSH{
  D.~Erkal and D.~Kutasov,
  ``a-Maximization, Global Symmetries and RG Flows,''
[arXiv:1007.2176 [hep-th]].
}
\lref\Kutasovii{
  D.~Kutasov and A.~Schwimmer,
  ``On duality in supersymmetric Yang-Mills theory,''
Phys.\ Lett.\ B {\bf 354}, 315 (1995).
[hep-th/9505004].
}
\lref\IntriligatorID{
  K.~A.~Intriligator and N.~Seiberg,
  ``Duality, monopoles, dyons, confinement and oblique confinement in supersymmetric SO(N(c)) gauge theories,''
Nucl.\ Phys.\ B {\bf 444}, 125 (1995).
[hep-th/9503179].
}
\lref\Kutasoviii{
  D.~Kutasov, A.~Schwimmer and N.~Seiberg,
  ``Chiral rings, singularity theory and electric - magnetic duality,''
Nucl.\ Phys.\ B {\bf 459}, 455 (1996).
[hep-th/9510222].
}
\lref\StrasslerQS{
  M.~J.~Strassler,
  ``The Duality cascade,''
[hep-th/0505153].
}
\lref\NSd{N.~Seiberg,
``Electric - magnetic duality in supersymmetric nonAbelian
gauge theories,''Nucl.\ Phys.\ B {\bf 435}, 129
(1995)[arXiv:hep-th/9411149].
}
\lref\PoppitzKZ{
  E.~Poppitz and M.~Unsal,
  ``Chiral gauge dynamics and dynamical supersymmetry breaking,''
JHEP {\bf 0907}, 060 (2009).
[arXiv:0905.0634 [hep-th]].
}
\lref\CsakiFM{
  C.~Csaki and H.~Murayama,
  ``New confining N=1 supersymmetric gauge theories,''
Phys.\ Rev.\ D {\bf 59}, 065001 (1999).
[hep-th/9810014].
}
\lref\BuicanEC{
  M.~Buican,
  ``Non-Perturbative Constraints on Light Sparticles from Properties of the RG Flow,''
[arXiv:1206.3033 [hep-th]].
}
\lref\DottiWN{
  G.~Dotti and A.~V.~Manohar,
  ``Supersymmetric gauge theories with an affine quantum moduli space,''
Phys.\ Rev.\ Lett.\  {\bf 80}, 2758 (1998).
[hep-th/9712010].
}
\lref\IntriligatorNE{
  K.~A.~Intriligator and P.~Pouliot,
  ``Exact superpotentials, quantum vacua and duality in supersymmetric SP(N(c)) gauge theories,''
Phys.\ Lett.\ B {\bf 353}, 471 (1995).
[hep-th/9505006].
}
\lref\Cardy{
J.~L.~Cardy,
``Is There A C Theorem In 4d?,''
Phys.\ Lett.\ B {\bf 215}, 749 (1988).
}
\lref\FreedmanWX{
  D.~Z.~Freedman, M.~Headrick and A.~Lawrence,
  ``On closed string tachyon dynamics,''
Phys.\ Rev.\ D {\bf 73}, 066015 (2006).
[hep-th/0510126].
}
\lref\KaplanKR{
  D.~B.~Kaplan, J.~-W.~Lee, D.~T.~Son and M.~A.~Stephanov,
  ``Conformality Lost,''
Phys.\ Rev.\ D {\bf 80}, 125005 (2009).
[arXiv:0905.4752 [hep-th]].
}
\lref\LutyWW{
  M.~A.~Luty, J.~Polchinski and R.~Rattazzi,
  ``The $a$-theorem and the Asymptotics of 4D Quantum Field Theory,''
[arXiv:1204.5221 [hep-th]].
}
\lref\ManoharIY{
  A.~V.~Manohar,
  ``Wess-Zumino terms in supersymmetric gauge theories,''
Phys.\ Rev.\ Lett.\  {\bf 81}, 1558 (1998).
[hep-th/9805144].
}
\lref\SeibergBZ{
  N.~Seiberg,
  ``Exact results on the space of vacua of four-dimensional SUSY gauge theories,''
Phys.\ Rev.\ D {\bf 49}, 6857 (1994).
[hep-th/9402044].
}
\lref\HofmanAR{
  D.~M.~Hofman and J.~Maldacena,
  ``Conformal collider physics: Energy and charge correlations,''
JHEP {\bf 0805}, 012 (2008).
[arXiv:0803.1467 [hep-th]].
}
\lref\osborn{
H.~Osborn,
``Derivation Of A Four-Dimensional C Theorem,''
Phys.\ Lett.\ B {\bf 222}, 97 (1989);
H.~Osborn,
``Weyl Consistency Conditions And A Local Renormalization Group Equation For
General Renormalizable Field Theories,''
Nucl.\ Phys.\ B {\bf 363}, 486 (1991);
I.~Jack and H.~Osborn,
``Analogs For The C Theorem For Four-Dimensional Renormalizable Field
Theories,''
Nucl.\ Phys.\ B {\bf 343}, 647 (1990).
}
\lref\HookFP{
  A.~Hook,
  ``A test for emergent dynamics,''
[arXiv:1204.4466 [hep-th]].
}
\lref\AharonyNE{
  O.~Aharony, J.~Sonnenschein and S.~Yankielowicz,
  ``Flows and duality symmetries in N=1 supersymmetric gauge theories,''
Nucl.\ Phys.\ B {\bf 449}, 509 (1995).
[hep-th/9504113].
}
 \lref\AEFJ{D.~Anselmi, J.~Erlich, D.~Z.~Freedman and A.~A.~Johansen,
``Positivity constraints on anomalies in supersymmetric gauge
theories,''
Phys.\ Rev.\ D {\bf 57}, 7570 (1998)
[arXiv:hep-th/9711035].
}
\lref\IntriligatorMI{
  K.~A.~Intriligator and B.~Wecht,
Nucl.\ Phys.\ B {\bf 677}, 223 (2004).
[hep-th/0309201].
}
\lref\AFGJ{D.~Anselmi, D.~Z.~Freedman, M.~T.~Grisaru and A.~A.~Johansen,
``Nonperturbative formulas for central functions of supersymmetric gauge
theories,''
Nucl.\ Phys.\ B {\bf 526}, 543 (1998)
[arXiv:hep-th/9708042].
}
\lref\DKlm{
D.~Kutasov,
 ``New results on the 'a-theorem' in four dimensional supersymmetric field
theory,''
arXiv:hep-th/0312098.
}
\lref\SchwimmerZA{
  A.~Schwimmer and S.~Theisen,
  ``Spontaneous Breaking of Conformal Invariance and Trace Anomaly Matching,''
Nucl.\ Phys.\ B {\bf 847}, 590 (2011).
[arXiv:1011.0696 [hep-th]].
}
\lref\KomargodskiVJ{
  Z.~Komargodski and A.~Schwimmer,
  ``On Renormalization Group Flows in Four Dimensions,''
JHEP {\bf 1112}, 099 (2011).
[arXiv:1107.3987 [hep-th]].
}
\lref\DongENA{
  X.~Dong, B.~Horn, E.~Silverstein and G.~Torroba,
  ``Unitarity bounds and RG flows in time dependent quantum field theory,''
[arXiv:1203.1680 [hep-th]].
}
\lref\IntriligatorMI{
  K.~A.~Intriligator and B.~Wecht,
  `RG fixed points and flows in SQCD with adjoints,''
Nucl.\ Phys.\ B {\bf 677}, 223 (2004).
[hep-th/0309201].
}
\lref\FortinCQ{
  J.~-F.~Fortin, B.~Grinstein and A.~Stergiou,
  ``Scale without Conformal Invariance in Four Dimensions,''
[arXiv:1206.2921 [hep-th]].
}
\lref\IntriligatorRX{
  K.~A.~Intriligator, N.~Seiberg and S.~H.~Shenker,
  ``Proposal for a simple model of dynamical SUSY breaking,''
Phys.\ Lett.\ B {\bf 342}, 152 (1995).
[hep-ph/9410203].
}
\lref\BrodieVV{
  J.~H.~Brodie, P.~L.~Cho and K.~A.~Intriligator,
  ``Misleading anomaly matchings?,''
Phys.\ Lett.\ B {\bf 429}, 319 (1998).
[hep-th/9802092].
}
\lref\KomargodskiXV{
  Z.~Komargodski,
  ``The Constraints of Conformal Symmetry on RG Flows,''
[arXiv:1112.4538 [hep-th]].
}
\lref\IntriligatorIF{
  K.~A.~Intriligator,
  ``IR free or interacting? A Proposed diagnostic,''
Nucl.\ Phys.\ B {\bf 730}, 239 (2005).
[hep-th/0509085].
}
\lref\KPS{
D.~Kutasov, A.~Parnachev and D.~A.~Sahakyan,
``Central charges and U(1)R symmetries in N = 1 super Yang-Mills,''
JHEP {\bf 0311}, 013 (2003)
[arXiv:hep-th/0308071].
}
\lref\KPS{
D.~Kutasov, A.~Parnachev and D.~A.~Sahakyan,
``Central charges and U(1)R symmetries in N = 1 super Yang-Mills,''
JHEP {\bf 0311}, 013 (2003)
[arXiv:hep-th/0308071].
}
\lref\VartanovXJ{
  G.~S.~Vartanov,
  ``On the ISS model of dynamical SUSY breaking,''
Phys.\ Lett.\ B {\bf 696}, 288 (2011).
[arXiv:1009.2153 [hep-th]].
}

\lref\KutasovFR{
  D.~Kutasov, J.~Lin and A.~Parnachev,
  ``Conformal Phase Transitions at Weak and Strong Coupling,''
Nucl.\ Phys.\ B {\bf 858}, 155 (2012).
[arXiv:1107.2324 [hep-th]].
}
\lref\FortinHN{
  J.~-F.~Fortin, B.~Grinstein and A.~Stergiou,
  ``A generalized c-theorem and the consistency of scale without conformal invariance,''
[arXiv:1208.3674 [hep-th]].
}
\lref\AmaritiSZ{
  A.~Amariti, L.~Girardello, A.~Mariotti and M.~Siani,
JHEP {\bf 1102}, 092 (2011).
[arXiv:1003.0523 [hep-th]].
}
\newbox\tmpbox\setbox\tmpbox\hbox{\abstractfont }
\Title{\vbox{\baselineskip12pt \hbox{UCSD-PTH-12-12}}}
{\vbox{\centerline{$\Delta a$ curiosities in some 4d susy RG flows}}}
\smallskip
\centerline{Antonio Amariti and Kenneth Intriligator}
\smallskip
\bigskip
\centerline{{\it Department of Physics, University of
California, San Diego, La Jolla, CA 92093 USA}}
\bigskip
\vskip 1cm

\noindent We explore some curiosities in 4d susy RG flows.   One issue is that the compelling candidate a-function, from a-maximization with Lagrange multipliers, has a `strange branch," with reversed RG flow properties,  monotonically increasing instead of decreasing.  The branch flip to the strange branch occurs where a  double-trace deformation $\Delta W=\CO ^2$ passes through marginality, reminiscent of the condition for the chiral symmetry breaking, out of the conformal window  transition  in non-susy gauge theories.  
The second issue arises from Higgsing vevs for IR-free fields, which sometimes superficially violate the a-theorem.  The resolution is that some vevs trigger marginal or irrelevant interactions,  leading to $\Delta a=0$ and decoupled dilaton on a subspace of the moduli space of vacua.   This  is contrary to classical intuition about Higgsing.  This phenomenon often (but not always) correlates with negative R-charge for the Higgsing chiral operator.

\bigskip

\Date{September 2012}

\newsec{Introduction}

Cardy's conjecture \Cardy\ is that the conformal anomaly $a$ counts the d.o.f. of 4d CFTs,  that all 4d RG flow endpoints satisfy $\Delta a\equiv a_{IR}-a_{UV}\leq 0$, and\foot{The fact that $a>0$ at the endpoints follows from \HofmanAR, since unitarity implies $c>0$.} $a>0$.   Intuitively then, $-\Delta a$ gives the net length of the RG flow.  There has been recent, renewed interest in the a-theorem, following the   recent, compelling arguments based on  $a$ anomaly matching \SchwimmerZA\ and associated unitarity constraints on the sign of $\Delta a$ \refs{\KomargodskiVJ, \KomargodskiXV}.      For spontaneous conformal symmetry breaking (or explicit soft breaking, via spurions), $\Delta a$ is related in   \KomargodskiVJ\ to the inclusive total scattering cross section $\sigma _{\tau \tau}(s)$ for the dilaton,  
\eqn\adiffKS{-\Delta a\equiv a_{UV}-a_{IR}={f^4\over \pi}\int _{s>0}ds{\sigma _{\tau \tau}(s)\over s^2}\geq 0.}
For exactly marginal deformations, there is no RG flow so $\Delta a=0$.  Intuitively, other, interacting deformation will lead to a RG flow and thus have $\sigma _{\tau \tau}(s)\neq 0$, and hence $\Delta a\neq 0$.  We will here discuss counter-intuitive examples, of Higgsing flows with $\Delta a=0$.

The statement that $\Delta a \leq 0$ for endpoints of RG flows is referred to as the ``weak version" of the a-theorem.   There is also the possibility of a stronger\foot{There is also the ``strongest" conjectured version of the a-theorem, that RG flow is gradient flow of $\tilde a(g)$ with positive definite metric on the space of deformations.  The validity of this conjecture, even in 2d conformal perturbation theory, awaits a better understanding of the appropriate metric on the space of deformations beyond lowest order \refs{\ZamolodchikovGT,\FreedmanWX}.  In 4d non-susy theories, there are recent works aiming towards perhaps producing counter-examples, e.g. \FortinCQ\ and references therein.}  version: that there exists a monotonically decreasing a-function $\tilde a(g)$, that is critical at the ends of RG flows where it reduces to $a$,  analogous to the  c-function in 2d \ZamolodchikovGT.  There is a 4d proposal \osborn\ that was checked in perturbation theory,  and explored more recently in e.g. \refs{\LutyWW, \FortinHN}.  In the susy context, there is another proposal \refs{\DKlm, \KSLM, \BarnesJJ}, that we'll further explore here.    

For  4d $N=1$ susy theories, results of \refs{\AEFJ, \AFGJ}\ relate $a$ at the endpoints of RG flows to superconformal $U(1)_{R_*}$ 't Hooft anomalies.  The $U(1)_{R_*}$ symmetry can often be exactly determined, using a-maximization \IW\ if needed.  This ``almost proved" the a-theorem in the supersymmetric context \IW.  The ``often" and ``almost" qualifiers are needed because of accidental symmetries \IW, which have crucial effects \KPS\ if present, see also e.g. \refs{\IntriligatorMI, \IntriligatorIF}.    

As observed in \DKlm, a-maximization of $a(R, \lambda)$, with $\lambda$ Lagrange multipliers for the interaction constraints on the superconformal R-charges $R$, gives a compelling candidate for the stronger versions of the a-theorem: with $a(\lambda)$ interpreted as an a-function along the RG flow, with $\lambda\sim g^2$ and with derivatives related to beta functions \refs{\DKlm, \KSLM, \BarnesJJ, \ErkalSH}.   

We here explore a few open issues \BarnesJJ. To briefly exhibit them, consider the function
\eqn\fis{  a_1(R)\equiv 3(R-1)^3-(R-1),}
which enters in a-maximization \IW.  As plotted in Fig. 1, $a_1(R)$ has a local maximum at the free-field value, $R={2\over 3}$, and a local minimum at $R={4\over 3}$.  Indeed $a_1(R)=-a_1(2-R)$, so $a_1(R=1)=0$, fitting with massive operators contributing $a=0$.     
 \bigskip
\centerline{\epsfxsize=0.60\hsize\epsfbox{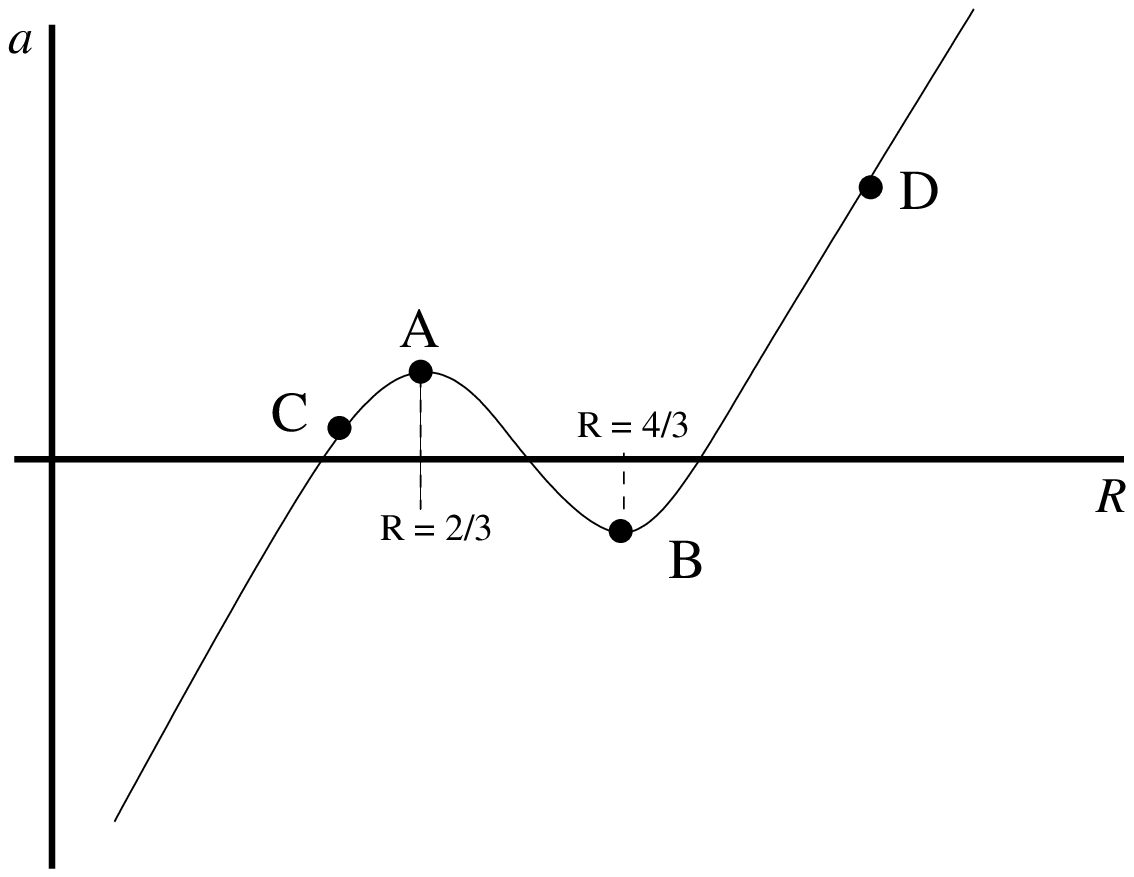}}
\centerline{\ninepoint\sl \baselineskip=2pt {\bf Figure 1:}
{\sl The function  $a_{1}(R)$.}}
\bigskip\noindent
One issue is a puzzling behavior of the conjectured a-function $a(\lambda)$ in RG flows past 
$R(\CO)=1$, which involves switching branches of a square-root, from one that is normal to one that is strange \BarnesJJ.    The flow direction has puzzling reversals on the strange-branch, with $a(\lambda)$ monotonically {\it increasing} rather than decreasing.  The endpoints of the flow nevertheless satisfy the weak version of the a-theorem, $\Delta a<0$.  

We note that the two branches give operators $\CO _\pm$ with a similar relation as that between an operator and its Legendre-transform source ``shadow" operator: the product $\CO _+\CO _-$ is a marginal superpotential term.  The branch flip occurs at the dividing line between where the ``double-trace" deformation $\Delta W=\CO _{cp}^2$ is relevant vs irrelevant.  In the non-susy context, the lore (see e.g. \refs{\CohenSQ,  \KaplanKR, \KutasovFR} and references cited therein) is that having a relevant $\Delta {\cal L}=\CO ^2$ triggers the phase change  from conformal to chiral symmetry breaking, $\vev{\CO}\neq 0$.   The susy case is different, but perhaps the phenomena are somehow related.

Another curiosity is how Higgsing is compatible with the a-theorem, particularly when the field with $\vev{Q}\neq 0$ has $R_{micro}(Q)\leq 0$.   Whenever a theory is Higgsed, the uneaten matter contributions {\it increase} $a$, so the a-theorem relies on having a greater contribution from the eaten matter, which moves from point $C$ in Fig. 2 to  $R_{IR}(Q_{eaten})=0$ \BarnesJJ.  This raises the question of what happens if initially point $C$ is at $R_{micro}(Q)\leq 0$, since then all contributions to $\Delta a$ are apparently contrary to the a-theorem.  It is crucial in these cases to account for the accidental symmetries.  We explore this issue further here.  Needless to say, we do not find a contradiction with the (weak version of the) a-theorem.

  \bigskip
\centerline{\epsfxsize=0.60\hsize\epsfbox{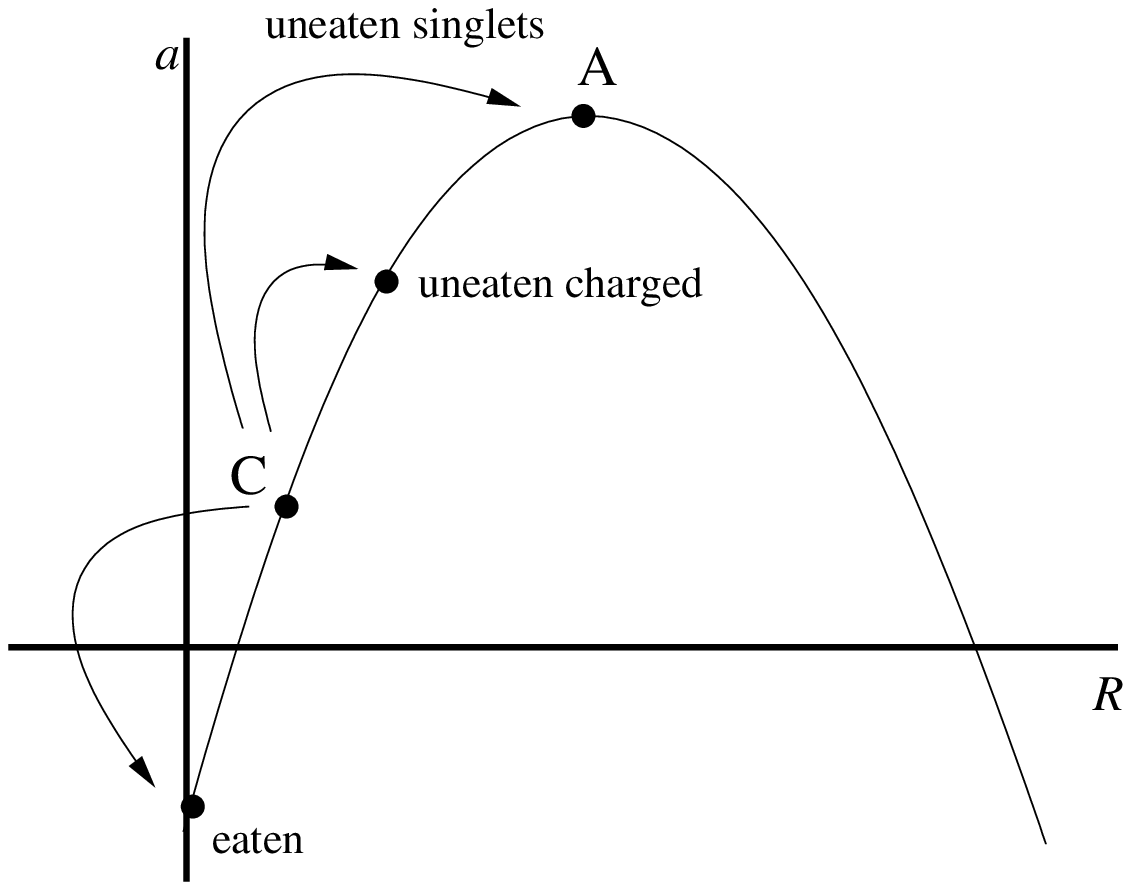}}
\centerline{\ninepoint\sl \baselineskip=2pt {\bf Figure 2:}
{\sl The change in $a$ from Higgsing.}}
\bigskip

It is interesting how the contradiction is avoided: the Higgsing can be marginal or trigger an irrelevant interaction, so $\Delta a=0$ in the end.   This does not happen for weakly coupled gauge theories.   Intuitively, Higgsing is always relevant, since operator vevs involve boundary conditions at long distance, and results in some fields getting massive, hence $\Delta a<0$.  
In SQCD, for example, moving away from the origin of the moduli space generally leads to $\Delta a<0$, whether the theory is in the conformal window or the IR free-magnetic phase.    This fits with the argument of \KomargodskiVJ\ that $-\Delta a$ is related as in \adiffKS\ to the total  dilaton scattering cross section.  In the examples that we explore, however, the operator Higgsing triggers an irrelevant or marginal interaction.  The endpoint of the (Wilsonian) flow has $\Delta a=0$, meaning that the dilaton is a completely decoupled free field, with $\sigma _{\tau \tau}(s)=0$.

An illustrative toy model has chiral superfields, $X$ and $\Phi$, with superpotential
\eqn\wtoy{W=h X \Phi ^n.}
For $n=1$, this is a mass term, while for $n\geq 2$ the coupling $h$ is irrelevant in the IR, $h\to 0$, 
so $X$ and $\Phi$ are IR free massless fields. Now consider deforming this theory by  $\vev{X}\neq 0$.    For $n=2$, \wtoy\ changes from being (marginally) irrelevant to relevant, giving $\Phi$ a mass 
$m_\Phi \sim h\vev{X}$, so the IR theory now only has the massless modulus $X$, and $\Delta a<0$.  For $n\geq 3$, on the other hand, \wtoy\ remains irrelevant even for $\vev{X}\neq 0$.  Since $\ev{X}\neq 0$ triggers an irrelevant interaction, an a-function would increase along this flow, with the final result that $\Delta a=0$ at the endpoint of the flow.   
 
We consider susy gauge theory examples similar to this toy model, where  $X$ can be a composite operator made up of charged matter constituents, $X=\prod _i Q_i^{p_i}$.  Then 
$\vev{X}\neq 0$ corresponds to $\vev{Q_i}=v_i \neq 0$, which Higgses the gauge group.    Classically, this gives gauge field and matter masses $\sim v_i$, so one would expect $\Delta a<0$.  Nevertheless, with IR free operators $X$, Higgsing $\vev{X}$ can trigger an interaction that can be either relevant or irrelevant, as in \wtoy.  In the latter case, $\Delta a=0$.    We note a frequent, though not strict, connection between this phenomenon and the sign of $R(Q)$.

\newsec{Review of susy results}

Supersymmetry relates the dilatation current to a particular $U(1)_R$ current, with 
\eqn\drreln{\Delta (\CO _{cp})={3\over 2}R(\CO _{cp})}
for scalar chiral primary operators.  This is interpreted as holding along the entire RG flow, even though the associated R-current is conserved only at the SCFT endpoints.    Although only  gauge invariant operators are observables, it is useful  to assign R-charges to the microscopic gauge fields of the underlying lagrangian (if it exists).  If the gauge invariant operator is $X=\prod _i Q_i^{p_i}$, then
\eqn\rmicro{R_{micro}(X)\equiv \sum _i p_i R(Q_i),}
and $R(X)=R_{micro}(X)$ unless $X$ is an IR-free field:
\eqn\rmicro{R(X)=\cases{ R_{micro}(X) & interacting $X$\cr {2\over 3} & IR free field $X$.}}
In the IR free case, the $U(1)_R$ mixes with an accidental $U(1)_X$ symmetry, which acts only on the IR free field composite $X$.  Having $R_{micro}(X)<2/3$ is a sufficient (by unitarity) condition that $X$ must be an IR-free field, though not necessary. See e.g.  \refs{\IntriligatorIF,\BuicanTY, \BuicanEC } for proposed diagnostics for IR free operators. 

The exact superconformal R-symmetry locally maximizes \IW\ 
\eqn\aris{a(R)=3\Tr R^3-\Tr R,}
with conformal anomaly at the RG fixed point given\foot{Note the normalization $a_{\rm here}={32\over 3}a_{\rm usual}$, so $a(\hbox{free chiral})={2\over  9}$ and $a(\hbox{free vector})=2$} by $a=a(R_*)$.  If there are accidental symmetries associated with IR free field operators $X$, the superconformal R-symmetry maximizes a modified function \KPS
\eqn\aacc{a_{X=free}(R)=a_{old}(R)+{\rm dim}(X)g(R_X),}
where ${\rm dim}(X)$ is the number of $X$ operators, and $g(R)$ is given in terms of  \fis\ by 
\eqn\gis{g(R)\equiv a_1(2/3)-a_1(R)= {1\over 9} (2-3R)^2(5-3R).}

It was conjectured in \DKlm\ that a-maximization can be extended to the RG flow via 
\eqn\arlam{a(R, \lambda)=a(R)+\lambda _I\widehat \beta ^I(R) \to a(\lambda)\equiv a(R(\lambda), \lambda),}
with $\lambda _I$ Lagrange multipliers and $\widehat \beta ^I(R)$ linear functions of $R$ that are proportional to the beta functions \refs{\DKlm, \KSLM, \BarnesJJ, \ErkalSH}.  Here $a(R)$ is given by \aris, or if there are accidental symmetries by \aacc\ (patched to \aris\ where the accidental symmetry arises on the flow).  The $\to$ in \arlam\ involves determining $R(\lambda)$ by extremizing $a(R, \lambda)$.     

The conjectured interpretation is that the  $\lambda ^I$ are related to the associated coupling constants,  $\lambda ^I\sim (g^I)^2$.   The result for $R(\lambda)$ then gives the exact anomalous dimension $\gamma _i$ of microscopic Lagrangian fields $Q_i$ in terms of their one-loop anomalous dimensions $\gamma ^{(1)}(g)$: 
\eqn\anomds{\gamma _{i\pm} =3R_\pm (Q_i)-2=1\mp \sqrt{1-2\gamma ^{(1)}(g)},}
with $\gamma ^{(1)}(g)$ linear in the $\lambda ^I$, and our notation is that $\gamma_{i+}$ is the normal branch, which connects with perturbation theory.  The $\gamma _{i-}$ solution in \anomds\  is the strange branch, that we'll discuss further.   By construction, $a(\lambda)$ in \arlam\ has 
\eqn\alamd{{\partial a(\lambda)\over\partial \lambda _I}=\widehat \beta ^I=f^I_J (g) \beta ^J(g).}
This suggests the possibility of RG gradient flow \refs{\DKlm, \BarnesJJ, \KSLM}
\eqn\agder{{\partial a\over \partial g^I}= G_{IJ}(g)\beta ^J (g), \qquad \hbox{with}\qquad G_{IJ}(g)\equiv f^K_J(g){\partial \lambda _K\over \partial g^I},} {\bf if} it turns out that $G_{IJ}(g)$ satisfies the conditions $G_{IJ}(g)>0$ and $G_{IJ}=G_{JI}$.

Near weak coupling, $\lambda \ll 1$ on the normal branch, \anomds\ can be matched to perturbation theory, relating  $\lambda _I$ to $g^I$, and matching \anomds\ to perturbation theory 
 \refs{\DKlm, \KSLM, \BarnesJJ}; moreover, $a(\lambda)$ and $G_{IJ}(g)$ in \agder\ nicely agree \BarnesJJ\ with Osborn's perturbative expressions \FreedmanRD.   The Lagrange-multiplier method can just as easily be applied to analyze deformations of a strongly coupled or non-Lagrangian initial SCFT, where again $\lambda \ll 1$ can be successfully compared with conformal perturbation theory around the UV SCFT \ErkalSH.  

\newsec{Cases where $a(\lambda)$  isn't monotonically decreasing?}

Extremizing the cubic function $a(R,\lambda)$ in $R$  in \arlam\  has two branches of solutions, $R_\pm (\lambda)$.  The $R_+(\lambda)$ solution is the normal branch, which connects with perturbation theory around the UV SCFT, while $R_-(\lambda)$ is the strange branch.  $R_+(\lambda)$ locally maximizes $a(R, \lambda)$, while $R_-$ is a local minimum.    Some RG flows go from the normal branch in the UV to the strange branch in the IR.  A simple example where the flip to the strange branch is needed is the magnetic dual of SQCD, when the dual quarks have $R(q)<\half$ \BarnesJJ.  The strange branch is generally needed if some field has sufficiently large, positive  anomalous dimension.  In the simplest examples, there is an operator $S$ such that the two branches have $R_+(S)+R_-(S)=2$ and the branch flip occurs at $R_+=R_-=1$.

Recall the notion of ``shadow fields" where, for each operator $\CO$ of dimension $\Delta$, one formally introduces an operator $\tilde \CO$ of dimension $d-\Delta$, whose one-point function can play the role of a source term for $\CO$.  This is familiar from AdS/CFT, where the two branches in solving for $\Delta _\pm (m)$ can be regarded as dimensions of conjugate operators $\Delta _\pm =\Delta (\CO_\pm)$.  With chiral superfields, it is natural to have shadow chiral superfields that are conjugate in terms of superpotential couplings, i.e. $\CO _\pm$ with $R(\CO _+)+R(\CO _-)=2$. As we discuss, the branch flip to the strange branch corresponds to an exchange in the roles of the operators and their conjugate shadow fields $\CO _\pm \leftrightarrow \CO _\mp$, and it occurs at $\CO _+=\CO _-=\CO$, where the double trace deformation $\Delta W=\CO ^2$ is marginal.    

To illustrate the strange branch phenomenon in a general context, consider deforming an initial SCFT by coupling some of its operators $\CO$ to some additional fields, $S$:
\eqn\ws{W=hS \CO,}
with $h$ the coupling constant.  The UV limit of the flow has $h\to 0$, $R_{UV}(S)={2\over 3}$, 
and $R_{UV}(\CO)<4/3$ for \ws\ to be relevant in the IR.    Generally, the RG flow from \ws\ affects both $R_S\equiv R(S)$ and $R_\CO\equiv R(\CO)$, but for simplicity we first consider cases where  
only $R_S$ flows, with $R_\CO $ fixed.  This is the case e.g. in the magnetic dual of SQCD, where $S\to M$ and $\CO \to q\tilde q$'s, with $R_\CO$ fixed, independent of $h$, by the symmetry. 

In this fixed $R_\CO$ case, the  Lagrange multiplier description \arlam\ gives 
\eqn\arex{a(R_S, \lambda)=a_{UV}+N_S (a_1(R_S) - {2\over 9})+\lambda (R_S+R_\CO -2),}
where $a_{UV}$ is the total $a$ of the UV SCFT (including the UV-free $S$ contributions), 
$N_S$ is the number of $S$ fields ($N_s=N_f^2$ in the SQCD example of \BarnesJJ), and $a_1(R_S)$ is the cubic function  \fis.   Extremizing the cubic function \arex\ in $R_S\equiv R(S)$ gives two solutions
\eqn\ars{R_S(\lambda, \epsilon _S)=1-{\epsilon _S\over 3}\sqrt{1-{\lambda \over N_S}},}
with $\epsilon_S=\pm 1$ labeling the two branches, with $\epsilon _S=+1$ the normal branch, and $\epsilon _S=-1$ the strange branch.   The normal branch has ${2\over 3}\leq R_S\leq 1$, with $0\leq \lambda \leq \lambda _{\max}=N_S$, and includes conformal perturbation theory around the the UV limit of the RG flow.  The strange branch is needed for $R_S>1$, and has $1\leq R_S\leq {4\over 3}$ ($R\leq 4/3$ was also emphasized recently in \HookFP), corresponding to $\lambda _{max}\geq \lambda \geq 0$ \BarnesJJ.    

If $R_\CO \geq 1$, then the theory stays on the normal branch, with no puzzling behavior.  If, on the other hand, $R_\CO<1$, then eventually $R_S$ flows to $R_S>1$, which requires going to the strange branch. In this process, $\lambda$ initially increases from $\lambda =0$ to $\lambda =\lambda _{\max}$ on the normal branch, and thereafter its flow direction must reverse, while on the strange branch, for $R_S$ to continue to increase past $R_S=1$.      As an extreme example, if $R_\CO =2/3$, then the theory flows first on the entire normal branch, from $\lambda =0$ in the UV to $\lambda =\lambda _{\rm max}$ in the middle of the flow, and then the flow continues across the strange branch, back to $\lambda =0$ in the extreme IR. There are two strange aspects of the strange branch: the $\lambda$ flow direction reverses, and $a(\lambda)$ increases rather than decreases along this flow.  

For fixed $\lambda$,  the normal branch locally maximizes $a(R,\lambda)$, whereas the strange branch locally minimizes it.   Indeed, plugging \ars\ back into \arex\ gives
\eqn\arexii{a(\lambda, \epsilon _S)=a_{UV}+{2\over 9}N_S \left[ \epsilon _S \left(1-{\lambda \over N_S}\right)^{3/2}-1\right]+ \lambda (R_\CO -1).}
So, for fixed $\lambda$, the value of $a$ is lower on the strange branch, $a(\lambda, +)>a(\lambda, -)$.  
Note that 
\eqn\arexd{{da \over d\lambda}=R_S+R_\CO -2=R_\CO -1-{\epsilon _S\over 3}\left(1-{\lambda \over N_S}\right)^{1/2}=\widehat \beta _h \leq 0,}
with a zero only at the endpoint $\lambda =\lambda _*$ of the RG flow, $da/d\lambda |_{\lambda =\lambda _*}=0$.  

The sign \arexd\ of the slope remains negative on both branches.  On the normal branch, the RG flow has $d\lambda >0$ and $da<0$, whereas on the strange branch $d\lambda <0$ and $da>0$, so $a(\lambda)$ monotonically decreases as expected on the normal branch, but monotonically increases with the RG flow on the strange branch.  Note also that
\eqn\arexdd{{d^2a\over d\lambda ^2}={1\over 6} {\epsilon _S\over N_S}\left(1-{\lambda \over N_S}\right)^{-1/2}}
is everywhere positive on the normal branch and everywhere negative on the strange branch, changing sign through $d^2a/d\lambda ^2=\infty$ at  $\lambda_{max}$.  
The branch flip point $\lambda _{\rm max}$ is the global minimum of $a(\lambda, \epsilon _S)$, even though \arexd\ is non-vanishing there.    On the strange branch, $a(\lambda)$ monotonically {\it increases},  up to a local {\it maximum} rather than minimum.  
 
Even though $a(\lambda)$ is not monotonically decreasing, the net change $\Delta a=a_{IR}-a_{UV}$ between the flow's endpoints is consistent with the weak version of the a-theorem, regardless of  $\lambda _*$'s branch: 
\eqn\deltaais{\Delta a= a(\lambda _*, \epsilon _S)-a_{UV}=-N_S g(R_S)<0.}
Here $g(R_S)$ is given in \gis, and $g(R_S)>0$ since $R_S<4/3$ ($g(R)>0$ for $R<5/3$).   
In particular, at weak coupling $R_S\approx {2\over 3}+{1\over 3}\gamma _S$, and $g(R_S)\approx {1\over 3}\gamma _S^2$.

Since $a(\lambda)$ is not monotonically decreasing, there are two options: (i) $a(\lambda)$ is not a good a-function after all, and needs modification; or (ii)  $a(\lambda)$ can be salvaged as an a-function, with some reinterpretation.
Regarding possibility (i), note that the $a_{UV}$ and $a_{IR}$ endpoints of $a(\lambda)$ are certainly correct, and we have found no associated contradiction with the weakest version of the a-theorem, $a_{IR}<a_{UV}$ for the RG flow endpoints.  So any modification cannot alter the RG flow endpoints.  This allows, for example, additive modifications proportional to powers of the beta-functions $a(g)\to a(g)+f(g) \beta (g)^p$.    Because $R(\lambda)$ and $a(\lambda)$ satisfy compelling and nontrivial checks with perturbation theory on the normal branch \refs{\DKlm, \KSLM, \BarnesJJ}, we could try to arrange for such modifications to kick in only at high order in perturbation theory, to try to preserve the good aspects on $a(\lambda)$ on the normal branch while fixing the puzzling aspects on the strange branch, outside of the perturbative regime.  Such a modification might work, but we have not found any naturally compelling candidate, nor a plausible reinterpretation for option (ii).   

A related question is the qualitative shape of the function $\lambda (h)$, which evidently cannot be one-to-one with the branch flip.  For example, $\lambda =0$ gives both $R_S=2/3$, hence $h=0$, on the normal branch, and $R_S=4/3$, hence $h\neq 0$ strongly coupled, on the strange branch. 
 The suggestion in \BarnesJJ\ is that $\lambda (h)$ has a ``shark-fin" shape, 
with $d\lambda (h)/dh^2>0$, in order for the Jacobian factor in the metric $G_{hh}$ in \agder\ to remain always positive.      

The shark fin shape, however, does not really help with the puzzle.   It is bad enough that the $\lambda$ flow direction reverses on the strange branch.  If $d\lambda /dh ^2$ remains positive, then the $h$ RG flow direction would also need to flip, switching from  $\dot h>0$ for the flow on the normal branch, to  $\dot h<0$ on the strange branch.  Such a flow direction reversal for a physical coupling would be unphysical: since RG flows are first order, $\dot h = -\beta (h)$, if $\beta (h)$ is continuous it would have to flow through a zero,  $\dot h = \beta (h)=0$, where the flow stops.

An alternative possible qualitative shape for $\lambda (h)$ is to have $d\lambda /dh^2<0$ after the branch flip, e.g. $\lambda (h) =  \lambda _{\rm max} \sin (\pi h^2/2 h_c^2)$, with branch flip at $h=h_c$, $\lambda =\lambda _{\rm max}$.  The advantage is that $h$ can continue increasing after the branch flip, where $\lambda$ decreases, consistent with $\beta (h)\neq 0$ away from the flow endpoints.  The metric $G_{hh}= f(h) d\lambda /dh$ \agder\ could then only be positive if $f(h)=\widehat \beta (R)/\beta (h)$ changes sign at $h=h_c$.  Since $\widehat \beta (R) \neq 0$ until the endpoint,  $f(h_c)=0$ requires $\beta (h_c)=\infty$, (e.g. analogous to the pole of the NSVZ beta function \ShifmanZI\ at $T(G)\alpha/2\pi =1$).  This merits further study.   But neither shape seems to change the fact that $a(\lambda)$ is not a monotonically decreasing function on the RG flow: it hits a minimum at the branch flip location and thereafter increases.

The example given in \BarnesJJ\ is the magnetic dual of SQCD \NSd, here with some renaming:  $SU(N_c)$ SQCD with $N_f^2$ added singlets, $S_{i\tilde j}$, coupled as in \ws\ to all the mesons $\CO _{i\tilde j}=M_{i\tilde j}=Q_i\tilde Q_{\tilde j}$.    The symmetries determine $R(Q)=(N_f-N_c)/N_f$, so $R(\CO)=R(M)=2(N_f-N_c)/N_f$, along the entire RG flow in $h$.  The branch flip to the strange branch is needed if $R(M)<1$, i.e. $N_f<2N_c$.  The extreme case is the bottom of the conformal window, where $R(\CO)\to {2\over 3}^+$, and $R(S)\to {4\over 3}^-$ in the IR, with $\lambda\to 0$ in the IR, on the strange branch.  Even though $a(\lambda)$ is not monotonic, the endpoints satisfy $\Delta a<0$ \deltaais. 

Let us now discuss the branches phenomenon in the completely general formulation of \ErkalSH\ of a-maximization with Lagrange multipliers.  The UV limit of the flow is an interacting SCFT $\CP$,  deformed by superpotential deformations $\Delta W=\sum _\alpha g^\alpha \CO _\alpha$ and/or gauging a subgroup of the global flavor group.   Along the resulting flow,  the superconformal R-current mixes with the remaining (ungauged) global symmetry currents $J^\mu _a$:
\eqn\rmix{R=R_\CP + d^a Q_a,}
where $d_a$ are real parameters that are determined by maximizing $a(\lambda, d)$ w.r.t. $d^a$:
\eqn\deqn{{\partial a\over \partial d^a}=0=-6\tau _{ab}d^b+9D_{abc}d^b d^c +\Lambda _a.}
See \ErkalSH\ for details and notation and, for convenience, we here define $\Lambda _a\equiv \lambda ^\alpha \bar\lambda _\beta (T_a)_\alpha ^\beta -\lambda _G k_a$.   The $\tau _{ab}$ in \deqn\ are the coefficients of the global current two-point functions, and $D_{abc}$ are the 't Hooft anomaly coefficients, $D_{abc}=\tr Q_aQ_bQ_c$,  both evaluated in the UV SCFT $\CP$.   We can choose the sign of the $Q_a$ charges such that all $D_{aaa}>0$.  In conformal perturbation theory near $\CP$, for small $\Lambda ^a$, the solution of \deqn\ is $d^a\approx {1\over 6}(\tau ^{-1})^{ab} \Lambda _b\equiv  {1\over 6}\Lambda ^a$,  on the normal-branch solution of \deqn.  

The strange branch is needed for sufficiently large $d^a$.  For example, for mixing with a single global current (so we can drop the $a,b,c$ indices) the solution of \deqn\ is 
\eqn\dsoln{d (\Lambda) ={\tau \over 3D}\left(1-\epsilon \sqrt{1-\Lambda D \tau ^{-2}}\right),}
with $\epsilon =+1$ the normal branch, and $\epsilon =-1$ the strange branch, with the branch flip at
\eqn\dflip{d_{\rm flip}=d(\Lambda _{\rm max})={\tau \over 3D}, \qquad 
\Lambda = \Lambda _{\rm max}=\tau ^2/D.}
Note also that $d$ has maximum value,  $d_{\rm max}=2d_{\rm flip}$, which is achieved on the strange branch, $\epsilon =-1$,  at $\Lambda =0$.    In terms of the R-charge \rmix, the branch flip occurs at
\eqn\rflip{R_{\rm flip}=R_\CP + {\tau \over 3D} Q.}  Taking $Q$ to act on the operators as $Q\CO _\alpha = T_\alpha ^\beta \CO _\beta$, the branch flip happens when the anomalous dimensions of operators change by a sufficiently positive amount, given by 
\eqn\gamflip{(\gamma _{\rm flip} )_\alpha ^\beta = (\gamma _\CP)_\alpha ^\beta +{\tau \over D}T_\alpha ^\beta.}
The case \ars\ is mixing with $U(1)_S$, with $Q(S)=S$, so $\tau =D=1$, and \dsoln\ reduces to agree with \ars\ and following expressions.

\newsec{``Double trace" deformations}

In the simplest examples \ws, the branch flip happens at $R(\CO)=1$, which is also where the double-trace operator $\Delta W\sim \CO ^2$ crosses through marginality.  Including the field $S$, the flow 
from the normal to the strange branch happens when $\Delta W=m_SS^2$ crosses from being relevant to irrelevant.  Perhaps this could help resolve the branch flip puzzles. 

As we discuss in this section, there is a close connection between the theory \ws\ with added singlets, and the double trace deformations.  Indeed, the singlets with coupling \ws\ can emerge dynamically from the double-trace deformations of an initial SCFT by 
\eqn\wcouple{\Delta W=h \CO ^2.}
We can also consider the product of two different operators,
\eqn\wcoupled{\Delta W=h\CO _1\CO _2,}
where  the operators $\CO _{1,2}$ can even be operators in two initially  decoupled theories, (SCFT)${}_1$ and $(SCFT)_2$ (see also e.g.  \DimoftePD).   To give a concrete example, we can consider SQCD and \wcouple\ could be the square of some meson operators, or \wcoupled\ could be meson operators in two different copies of SQCD.  If $R(\CO)<1$ (or $R(\CO_1)+R(\CO _2)<2$ for \wcoupled) in the initial SCFT, then \wcouple\ is relevant\foot{If $R(\CO)=1$, one can determine if $\Delta W$ is exactly marginal or marginally irrelevant \refs{\GreenDA, \ErkalSH}.} and drives the theory somewhere new.

The correct way to analyze the RG flow is via the usual trick of introducing some new fields $S$, which yields the double-trace deformation \wcouple\ upon integrating out $S$
\eqn\wcouples{W=h_S S \CO + \half m_S S^2.}
Likewise, we get \wcoupled\ by adding massive fields $S_1$ and $S_2$, with 
\eqn\wcoupleds{W=h_1S_1 \CO _1+h_2 S_2\CO _2+m_S S_1S_2}
The fields $S$ in \wcouples\ and \wcoupleds\ start as free fields, $R_{UV}(S)=2/3$ for $h=0$, where the $m_S$ terms are relevant. But one  should first analyze the RG flow $h\to h_*$ with $m_S=0$, 
\eqn\wint{W_{int}=h_S S\CO,}
 and subsequently include the $m_S\neq 0$ term, as a deformation of that theory.  This  
captures the fact that the $m_S$ terms kick in later in the RG flow to the IR.  

Let $R_0(\CO)\equiv R_{UV}(\CO)$ be the superconformal R-charge at the UV starting point, with all couplings in \wcouples\ or \wcoupleds\ set to zero, and $R_{1}(\CO)$ denote the superconformal R-charge at the IR endpoint of \wint\ ($h_S=h_*, m_S=0)$, and $R_{2}(\CO)$ denote that at the IR endpoint of  \wcouples\ or \wcoupleds\ ($h_S\neq 0$, $m_S\neq 0)$.  
   Since we suppose \wcouple\ is relevant, $R_0(\CO)<1$, which ensures that the $h$ term in \wcouples\ is relevant (that only requires $R_0<4/3$).  The $m_S$ coupling in \wcouples\ is irrelevant if $R_{1}(S)>1$, which means $R_{1}(\CO)<1$, and which is often the case. For example, if $R_0(\CO)=R_{1}(\CO)$, the condition  needed for \wcouple\ to be relevant also implies that the $m_S$ term in \wcouples\ is irrelevant, $m_S\to 0$.  

When the $m_S$ term is irrelevant, the upshot of the double trace deformation \wcouple\ is an emergent new field,  $S$, with superpotential \wint\ so $R(S)=2-R(\CO)$.  The $F_S$ equation of \wint\ removes $\CO$ from the chiral ring.  So the upshot is to replace $\CO \to S$, with $R(\CO)$ replaced with that of $S$, $R(S)=2-R(\CO)$.  This effect has already been discussed in the literature, in the context of many specific examples -- see e.g. \StrasslerQS\ for SQCD with the meson-squared deformation.  In the context of ADS/CFT, the bulk field with mass $m$ maps to operators $\CO _{\pm}$ with $\Delta _++\Delta _-=4$, so one is naturally an operator in the CFT and the other is it's shadow operator.  Then the flow of $h\CO _-^2$ and associated IR replacement $\CO _-\to \CO _+$ is again well known, see e.g. \DongENA. 

To connect with the previous section, note that a double trace deformation leads to the emergent $S$ field, with $m_S\to 0$ in the IR, precisely when the flow takes $R(S)>1$.  As in \ars, this coincides with where the RG flow flips to the  $\epsilon _S=-1$ branch.

\newsec{Examples: SQCDM${}^2$}

In this section we discuss a class of examples that are of interest, both because they illustrate double trace deformations, and for application to later sections, where we consider examples of $\ev{Q}\neq 0$ Higgsing of fields with $R_{micro}(Q)<0$.

Consider $SU(N_c)$ SQCD with $N_f+N_f'$ fundamental flavors, $N_f'$ of which enter in 
\eqn\sqcdw{W_{tree}= h^{i' \tilde i', j'\tilde j'} (Q'_{i'}\widetilde Q' _{\tilde i'})( Q'_{j'}\tilde Q'_{\tilde j'}).}
We refer to this theory as SQCD$M^2$.    The case where all flavors enter in \sqcdw, i.e. $N_f=0$, was considered in earlier works, e.g. \LeighEP.  In general, \sqcdw\ is relevant for $N_f^{tot}=N_f+N_f'<2N_c$, in which case \sqcdw\ drives a RG flow to a new IR fixed point.  Let us define 
\eqn\ssqcdd{R(Q)=R(\tilde Q)\equiv y, \qquad x\equiv N_c/N_f, \qquad n\equiv N_f'/N_f.}
Naively, the IR fixed point theory has $R(Q')=R(\tilde Q')=\half$ for the $N_f'$ fields in \sqcdw, since $R(W)=2$, and then $U(1)_R$ conservation (anomaly free) gives for the remaining $N_f$ fields
\eqn\ynaive{y_{naive}= {n\over 2}+1-x.}
For later use, we note that there is a range of allowed $n$ and $x$ where $y_{naive}<0$, suggesting that these theories as candidates for exploring Higgsing by $R<0$ operators, in this case by $M=Q\tilde Q$.  However, as we now discuss, the correct treatment of the double trace deformation \sqcdw\ shows that this class of examples actually always has $R(Q)>0$.

\subsec{SQCD coupled to added singlets: SSQCD}

The intermediate theory, needed to analyze the double-trace theory \sqcdw, is the SSQCD theory of \BarnesJJ (see also e.g. \AmaritiSZ ):
 $SU(N_c)$ SQCD with $N_f+N_f'$ fundamental flavors, coupled to $N_f'^2$ singlets
\eqn\wssqcd{W=h \sum_{i',\tilde j'=1}^{N_f'} S^{i' j'}Q'_{i'}\tilde Q'_{\tilde j'}.}
For $N_f'=0$ it is SQCD and for $N_f=0$ it is a magnetic dual SQCD.  For $N_f$ and $N_f'$ both non-zero, both $R(S)$ and $R(Q')$ vary along the RG flow, and a-maximization is required.  
Using the definitions \ssqcdd, the R-charge conservation constraints determine
\eqn\rconstraints{R(Q')=R(\tilde Q')=1+{1-y-x\over n}, \qquad R(S)=2\left({x+y-1\over n}\right).}
The vacuum stability bound condition, needed to avoid $W_{dyn}$ or a deformed moduli space,  is $N_f+N_f'>N_c$, i.e. $x<1+n$.    a-maximization is then used to determine $y(x, n)$ \BarnesJJ.  

For $x\geq x_M(n)$, there is an accidental symmetry associated with $M=Q\tilde Q$ being IR free, with $x_M(n)$ determined by the unitarity bound for $M$: $y(x_M(n))=1/3$.  The SSQCD theory has a $SU(N_f-N_c)$ dual \BarnesJJ, with superpotential (here we add the $m_S$ term)
\eqn\wssqcdual{W_{dual}=SM'+M'q\tilde q+M q'\tilde q'+Pq'\tilde q+P'q\tilde q'+\half m_S S^2.}    Knowledge of the dual shows that there are additional accidental symmetries, that of the IR free magnetic phase, when $x\geq x_{FM}={2\over 3}(1+n)$.    For $n$ large enough, there is a range $x_M(n)<x<x_{FM}$ where $M$ has hit the unitarity bound and decoupled, but the rest of the theory remains interacting.  In this case, the $Mq'\tilde q'$ term in \wssqcdual\ becomes irrelevant.  a-maximization shows that $y$ remains everywhere positive in any case.   

Note that SSQCD has a moduli space where $\vev{S}\neq 0$.  For generic $\vev{S}$ we can integrate out the massive $Q'\tilde Q'$ matter fields, and the low-energy theory consists of $SU(N_c)$ SQCD with $N_f$ matter fields $Q$, $\tilde Q$.  If $N_f<N_c$, that theory can generate a superpotential,
\eqn\wlowssqcd{W_{dyn}=(N_c-N_f)\left({\det S \Lambda ^{3N_c-N_f-N_f'}\over \det M}\right)^{1/(N_c-N_f)}.}

\subsec{Back to the double-trace theories SQCD$M^2$ }

The double trace theory \sqcdw\ is obtained as  a $m_S$ deformed version of the SSQCD theory discussed in \BarnesJJ\ and the previous subsection:
\eqn\ssqcdm{W_{elec}= S^{i'\tilde j'}Q'_{i'}\tilde Q'_{\tilde j'}+\half (m_S) _{i'\tilde i', j \tilde j'}S^{i\tilde i'}S^{j'\tilde j'},}
as in \wcouples.   We can now analyze whether the $m_S$ term is a relevant or irrelevant deformation of the SSQCD fixed point.  If $m_S$ is relevant, the IR limit of SQCD$M^2$ has massive $S^{ij}$ fields, with $R(S)=R(Q')=R(\tilde Q')=1$, and $R(Q)$ given by \ynaive.  If $m_S$ is irrelevant, the IR limit of SQCD$M^2$ is the same as that of SSQCD, with emergent $S^{ij}$ fields.

For the $N_f=0$ case  \refs{\LeighEP, \StrasslerQS}, the flavor symmetry fixes $R(Q'\tilde Q')=1-N_c/N_f'$ and then the condition needed for the double-trace term to be relevant, $N_f'<2N_c$, always implies that the $m_S$ term in \ssqcdm\ is irrelevant.  So this case always has  the emergent $S$ fields \ssqcdm, with $m_S\to 0$.    In the magnetic dual description, the original quartic superpotential \sqcdw\ is a mass term for the magnetic mesons $M'\sim Q'\tilde Q'$, which can be integrated out leading to a quartic superpotential in the dual, $W_{dual}= \tilde h(q\tilde q)^2$, which is IR irrelevant, $\tilde h\to 0$.  Then $q\tilde q\to S$ emerge from the magnetic dual in the IR as new chiral primaries.

For general $N_f$ and $N_f'$, the $m_S$ term in \ssqcdm\ is relevant if $R(S)<1$, i.e. if a-maximization in SSQCD yields $y(x, n)<1+{n\over 2}-x$, and $m_S$ is irrelevant otherwise.  When $m_S$ is relevant, its effect is to replace  $y_{SSQCD}\to 1+{n\over 2}-x$, as in \ynaive,  so in the IR
\eqn\ybetter{y_{SQCDM^2}(x, n)={\rm max}(y_{SSQCD}(x, n), 1+{n\over 2}-x).}
Using a-maximization it is found that $y_{SSQCD}(x, n)>0$ \BarnesJJ, so it follows from \ybetter\ that $y_{SSQCDM^2}(x, n)>0$, unlike \ynaive\ which can be negative, so $R_{micro}(M)>0$ in all case.   

In the later sections we will discuss examples where Higgsing leads to $\Delta a=0$.  Neither SSQCD nor SQCD$M^2$ are like that: in both theories, taking e.g. $\ev{M}\neq 0$, drives a relevant interaction, with $\Delta a<0$, even in the case where $M$ is IR free.   In the dual \wssqcdual, this  is because $R(q'\tilde q')<2$, so $\vev{M}$ always triggers a relevant deformation, leading to $\Delta a<0$.

\newsec{Higgsing $\vev{Q}$: when can it be irrelevant / marginal?}

Superperconformal theories generally have a moduli space of vacua, and one  can consider deforming the SCFT at the origin by giving expectation values to moduli fields, $\vev{X}=v\neq 0$, spontaneously breaking the conformal symmetry, with the modulus $X$ the massless dilaton. The beta functions vanish, but there can be a RG flow in a Wilsonian sense, with $v\to 0$ in the UV limit and the massive modes from $v\neq 0$ integrated out in the IR.   The IR limit has a restored, linearly realized conformal symmetry for the remaining massless fields.   The ``flow" has $a_{UV}=a_{origin}$ and $a_{IR}=a_{v\neq 0}$, with $-\Delta a \equiv a_{UV}-a_{IR}\geq 0$ related as in \adiffKS\ \KomargodskiVJ\ to the total scattering cross section $\sigma (s)$ for the modulus $X$.    

Intuitively, since $\ev{X}\neq 0$ changes the field boundary conditions at large distance, it should always be ``relevant," and non-trivially affect the theory in the IR.  This classical intuition suggests that vevs $\ev{X}\neq 0$ should lead to $\Delta a <0$, fitting with $\sigma (s)\neq 0$ in \adiffKS.  We will discuss examples where this classical intuition is wrong, and instead $\ev{X}$ triggers an irrelevant deformation.  Away from the RG flow endpoint, it should effectively leads to an {\it increase} of a hypothetical a-function.  In the deep IR, the irrelevant deformation relaxes away, and the endpoints satisfy $\Delta a=0$.  According to \adiffKS, $\Delta a=0$ means that $\sigma _{\tau \tau}(s)=0$, i.e. the modulus $\tau$ is a completely decoupled free field.   We refer to this as irrelevant / marginal Higgsing.  

Irrelevant / marginal Higgsing is indeed  unusual, e.g.  it does not happen in ${\cal N}=4$ or in ${\cal N}=1$ SQCD Higgsing.  In those cases, integrating out the massive fields, e.g. the W-bosons on the Coulomb branch of ${\cal N}=4$, leads to dilaton scattering derivative interactions (e.g. superpartners of the $F^4$ terms), and hence $\sigma _{\tau \tau} (s)\neq 0$ and $\Delta a\neq 0$.   Likewise, in SQCD, taking  $\vev{M}\neq 0$ is always a relevant deformation, leading to $\Delta a\neq 0$,  even in the free-magnetic range of $N_f$ \SeibergBZ, where the mesons $M$ are IR free.
If, for example, the gauge group is completely broken on the moduli space, the low-energy theory in the bulk of the moduli space ${\cal M}$ consists of the  IR-free moduli fields, so $a_{bulk}={2\over 9}{\rm dim}_C({\cal M})$.    The a-theorem implies that the theory at the origin has $a_{origin}\geq a_{bulk}$.  This is indeed what happens in e.g. SQCD for all  $N_f>N_c$. 

Again, from the perspective of Fig 2, the fact that $\Delta a\leq 0$ for Higgsing RG flows is non-obvious \BarnesJJ, particularly when the fields $Q$ with $\ev{Q}\neq 0$ have $R_{micro}(Q)<0$.    We will indeed see a frequent correlation between the irrelevant / marginal Higgsing phenomenon associated with $\ev{X}$ and negative R-charge, $R_{micro}(X)\leq 0$.   We can also motivate this connection in terms of gauge invariant operators, by considering 
\eqn\wlm{W=hL(X-X_{micro}),}
where $X_{micro}=\prod _i Q_i^{p_i}$, and $X$ and $L$ are added fields, with $R=2/3$ in the UV limit.  The $LX_{micro}$ term drives a RG flow to where $R(L)=2-R(X_{micro})$.  The $LX$ term in \wlm\ is relevant if $R(L)<{4\over 3}$, i.e. if  $R_{micro}(X)>2/3$; the added $L$ and $X$ fields are then massive and can be integrated out, $L\to 0$, $X\to X_{micro}$.  For  $R(X_{micro})\leq 2/3$, the $LX$ term in \wlm\ is irrelevant, so $X$ is an 
emergent, IR-free decoupled field  \BarnesJJ.  

For $0<R_{micro}(X) <{2\over 3}$ the $hLX$ interaction is dangerously irrelevant, in that it becomes relevant for $\vev{X}\neq 0$.    This simple argument suggests that $\ev{X}\neq 0$ will be different if $R_{micro}(X)\leq 0$, since then $R(L)>2$ and $W=hXL$ remains irrelevant for  $\ev{X}\neq 0$.    We note that irrelevant / marginal $\vev{X}$ indeed often correlates with $R_{micro}(X)<0$, but we note examples where that is not the case. In one class, operators with $R_{micro}(X)<0$ have relevant $\vev{X}$.  In another class, operators with $R_{micro}(X)>0$ have irrelevant $\vev{X}$.

Having operators with $R_{micro}(X)<0$ raises the possibility of a vacuum instability, having conformal symmetry broken by a dynamically generated superpotential, $W_{dyn}$.  We are here interested in theories with  $W_{dyn}=0$, despite having $R_{micro}(X)<0$ operators.

\newsec{ Some known, IR free examples of Higgsing with $\Delta a=0$ moduli spaces.}

The irrelevant / marginal Higgsing phenomenon can be found in a few examples already appearing in the literature (the $\Delta a=0$ curiosity was not noted or emphasized). 

  \subsec{$R_{micro}(X)=0$ examples with $\Delta a=0$, e.g. SQCD for $N_f=N_c$}
 
The matter fields here have $R_{micro}(X)=0$, compatible with the quantum modified constraint \SeibergBZ, which removes the origin from the moduli space.   The low-energy theory on the smooth moduli space is an IR-free theory of the constrained moduli fields $\CM$ (with a WZ term to account for 't Hooft anomaly matching of the global symmetries \ManoharIY), so $a_{IR}(\CM)={2\over 9} \dim _C\CM$.  Moving on the moduli space has $\Delta a=0$, so $\vev{X}$ can be regarded as ``marginal."  Different points on the moduli space can nevertheless  be physically distinguished, by their different massive spectrum and different IR unbroken global symmetries. 

This case can be contrasted with SQCD for $N_f\geq N_c+1$, where $\Delta a<0$ and $R(X)>0$. 
There the moduli space $\CM$
is singular, with additional massless states near the origin, giving $a_{origin}>a_{bulk}(\CM)$, where $a_{bulk}(\CM)={2\over 9}\dim _C\CM = {2\over 9}(2N_fN_c-(N_c^2-1))$.  For example, for $N_f=N_c+1$, $a_{bulk}(\CM)={2\over 9}N_f^2$ while the theory at the origin has $a_{origin}={2\over 9}(N_f^2+2N_f)>a_{bulk}$, thanks to the additional massless $B^i$ and $\tilde B^i$ fields at $M_{ij}=0$ \SeibergBZ.  

\subsec{$R_{micro}(X)<0$ affine moduli space examples with $\Delta a=0$}

There are examples where $R_{micro}(X)<0$, with $W_{dyn} =0$ thanks to a quantum-branch cancellation.  The original examples are  $SO(N_c)$ with $N_f=N_c-4$ fundamental flavors $Q_f$ \IntriligatorID: there are discrete quantum parameters, $\epsilon _{1,2}=\pm 1$, and $W_{dyn}=2(\epsilon _1+\epsilon _2)(\Lambda ^{2(N_c-1)}/\det M)^{1/2}$ vanishes on the $\epsilon _1=-\epsilon _2$ branches.  The $W_{dyn}=0$ branch has a smooth moduli space $\CM$, with no additional massless fields.  Indeed, the moduli saturate the 't Hooft anomalies.  Theories of this type, with simple gauge group and $W_{tree}=0$, were classified in \DottiWN.  The T1-T6 theories there have $\mu _{matter}<T(G)$, and  thus $R_{micro}(X)<0$ fields, and are all IR-free, with $a_{IR}(\CM)={2\over 9}\dim _C\CM$, constant on all of $\CM$.  In all these theories, the $W_{dyn}=0$ branch is lifted (broken susy) upon adding $W_{tree}\neq 0$. 

So Higgsing on $\CM$ has $\Delta a=0$: the moduli vevs $\ev{M}$ are marginal.  According to \adiffKS, the massless moduli are thus completely decoupled from any other fields,  $\sigma (s)=0$.  

\subsec{Examples with $\Delta a\neq 0$ Higgsing, fitting with $R_{micro}(X)>0$.}

Recall the example of $SU(2)$ with matter field $Q\in {\bf 4}$  \IntriligatorRX, with $R_{micro}(Q)=3/5$.  The $W_{tree}=0$ theory has a moduli space $\CM \cong {\bf C}$, with modulus $X=Q^4$.  Since $X$,  with $R_{micro}(X)=12/5$, saturates the $\Tr R$ and $\Tr R^3$ anomaly matching,  there are two scenarios \IntriligatorRX\ for the IR dynamics at $X=0$: (i): $X$ is a decoupled,  IR-free field (then $R_{IR}(X)=2/3$, via accidental symmetry), or (ii) it is a SCFT, with $R_{SCFT}(X)=12/5$ (with misleading anomaly matching \BrodieVV).  Several diagnostics \refs{\IntriligatorIF, \PoppitzKZ, \BuicanTY, \VartanovXJ} suggest that the correct IR phase is probably (ii),  SCFT (so $\Delta W=\lambda X$ is irrelevant, instead of susy-breaking). 

The IR free phase scenario (i) would have had $a_{IR free}={2\over 9}$ everywhere on $\CM$, so  $\Delta a=0$.   The presumably correct SCFT scenario (ii), on the other hand, has $a_{origin}= 3({7\over 5})^3-{7\over 5}$ (assuming that there are no overlooked accidental symmetries), and $a_{bulk}={2\over 9}$, so Higgsing by $\ev{X}$ has $\Delta a=a_{bulk}-a_{origin}<0$.  Assuming scenario (ii) is correct thus exhibits the frequent correlation between $R_{micro}(X)>0$ and $\Delta a\neq 0$.  More generally, the proposed diagnostic of \IntriligatorIF\ automatically ensures that Higgsing has  $\Delta a$ of correct sign.  

Similarly, the $SO(N_c)$ with symmetric tensor examples \BrodieVV, and the other $\mu _{matter}>T(G)$ examples classified in \DottiWN, all have $R_{micro}(X)>0$ moduli, and all lead to SCFTs at the origin, with $a_{SCFT}>a_{IR, free}$.  So Higgsing $\ev{X}\neq 0$ gives $\Delta a=a_{IR, free}-a_{SCFT}<0$, non-zero, again correlating $R_{micro}(X)>0$ with $\ev{X}$ ``relevant," leading to $\Delta a\neq 0$.

\subsec{$\Delta a=0$, $R_{micro}(X)<0$ examples, with higher power irrelevant interactions}

Consider the theory of \refs{\Kutasov, \Kutasovii, \Kutasoviii}, SQCD with adjoint $X$ and $W_{tree}=\Tr X^{k+1}$, choosing $N_f$ and $N_c$ such that $\tilde N_c\equiv kN_f-N_c=1$.  The magnetic ``$SU(1)$" group is then trivial, corresponding to a smoothly-confining theory, with 't Hooft anomalies matching between the original electric theory and an IR free dual ``confined" theory of generalized mesons and baryons.  Consider in particular the $k=2$ case,  $W=\Tr X^3$, with  $2N_f-N_c=1$.  The dual fields are the 
generalized mesons and baryons $M_1=(\tilde Q Q)$, $M_2=\tilde Q X Q$, $B= (Q^{N_f}(XQ)^{N_f-1})$, $\tilde B=(\tilde Q ^{N_f}(X\tilde Q)^{N_f-1})$, with \CsakiFM,
\eqn\wcm{W={1\over h^{2N_f-1}\Lambda ^{2N_f-4}}(\tilde B M_2 \tilde B-\det M_2(M_1{\rm cof}M_2)).}
The mesons have $R_{micro}(M_1)=2-4(2N_f-1)/3N_f<0$ and $R_{micro}(M_2)={8\over 3}-4(2N_f-1)/3N_f>0$.  For large $N_f$, $R_{micro}(M_1)\to -{2\over 3}$ and $R_{micro}(M_2)\to 0^+$.  Of course, since the dual theory is IR free, the physical values are $R(M_1)=R(M_2)=R(B)=R(\tilde B)=2/3$.  The values of $R_{micro}$ remain nevertheless useful to characterizing the irrelevant interactions in \wcm.  Clearly, $\vev{M_2}$ is relevant, since it gives a quadratic mass term to $B$ and $\tilde B$; this fits with $R_{micro}(M_2)>0$.  On the other hand, for $\vev{M_2}=0$, it is evident from \wcm\ that $\ev{M_1}$ has no effect; this is a marginal subspace of the moduli space, with $\Delta a=0$. This correlates with $R_{micro}(M_1)<0$, and is  analogous to the toy model \wtoy\ for $n>2$.

\subsec{$R_{micro}(X)<0$ counter-examples, with relevant $\vev{X}$ leading to $\Delta a\neq 0$.}

Recall the case of $SO(N_c)$ with $N_f=N_c-3$ \IntriligatorID: the mesons $M_{fg}=Q_f\cdot Q_g$ have $R_{micro}(M)<0$, and there is a branch with a quantum moduli space of the unconstrained mesons $M_{fg}$, with additional massless fields at the origin, $W_{low}=M_{ij}q^i q^j$.  Despite the fact that $R_{micro}(M)<0$, and thus $R_{micro}(q^2 )>2$, taking $\vev{M}\neq 0$ gives a mass deformation, which is always relevant, leading to $\Delta a<0$.  Of course, $R(M)=R(q)=2/3$, since the theory is IR free.  Even though $R_{micro}(M)<0$, the interactions are similar to the $n=2$ case of \wtoy\ and the above $0<R_{micro}(X)<2/3$ examples.

\subsec{$R_{micro}(X)>0$ counter-examples, with irrelevant $\vev{X}$ leading to $\Delta a=0$.} 

Recall the case of $Sp(N_c)$ with $N_f=N_c+2$ fundamental matter fields \IntriligatorNE.  There is a moduli space given by expectation values of $M_{ij}=Q_{ic}Q_{jd}J^{cd}$, in the $N_f(2N_f-1)$ antisymmetric rep of $SU(2N_f)$, with $R_{micro}(M)=2/N_f$ and superpotential interactions \eqn\wpf{W=-{{\rm Pf}\ M\over 2^{N_c-1}\Lambda ^{2N_c+1}},}
and $F_M=0$ gives the classical moduli space constraint $\half {\rm rank}(M)\leq N_c$.   
For the case $N_c=1$, $Sp(1)\cong SU(2)$, the superpotential \wpf\ is cubic, and $\vev{M} \neq 0$ 
gives a relevant mass terms for the non-classical, additional $M_{ij}$ fields that are massless at the origin, hence $\Delta a <0$.  This $Sp(1)$ case is in the $SU(N_c)$ with $N_f=N_c+1$ family\SeibergBZ, where $\ev{M_{ij}}\neq 0$ always gives a mass to some additional baryons, leading always to $\Delta a <0$.    

For higher $Sp(N_c)$,  on the other hand, the superpotential \wpf\ is order $N_f=N_c+2$ in the $M$ fields, so $\vev{M}$ of low enough rank can trigger an irrelevant interacting, leading to $\Delta a=0$ in the far IR.   This happens on a subspace of the moduli space, where $\ev{M}$ has $\half {\rm rank}(\ev{M})\leq N_c-1$, since then the interaction remaining from \wpf\ is cubic or higher order, so is irrelevant, giving $\Delta a=0$ on this noncompact subspace including the origin:
\eqn\asp{a (\vev{M}) =\cases{{2\over 9} N_f (2N_f-1) & for $\half {\rm rank}(M)\leq N_c-1$\cr {2\over 9}N_c(2N_c+7) & for  $\half {\rm rank}(M)=N_c.$\cr}}
The $\Delta a=0$ Higgsing phenomenon occurs on that subspace of the full moduli space where $Sp(N_c)$ is only partially Higgsed, with at least a $Sp(1)$ unbroken, where all $M_{ij}$ fields remain massless and IR free.  When $\half {\rm rank}(M)=N_c$, the $Sp(N_c)$ group is fully broken, and the non-classical $M_{ij}$ components have becomes massive; there the 
the $a$ value in \asp\ comes from simply counting the $Q$ matter fields left uneaten after the Higgs mechanism.

\newsec{Examples of relevant, $\Delta a\neq 0$ Higgsing for $R_{micro}(X)>0$ interacting SCFTs} 

Consider $SU(N_c)$ gauge theory with $N_f$ fundamentals, $Q_f$, $\tilde Q_{\tilde f}$ and $N_a$ adjoints, $X_i$, taking $W_{tree}=0$.  For $N_a=0$, the theory is SQCD; for $N_a=1$, the theory flows to the $\widehat A$ SCFTs  (for $1<N_f<2N_c$); for $N_a=2$, the theory flows to the $\widehat O$ SCFTs (for $N_f<N_c$).  See \refs{\KPS, \IntriligatorMI} for the detailed a-maximization analysis of these latter SCFTs.   As there, it is convenient to take the Veneziano limit of large $N_c$ and $N_f$, holding $x\equiv N_c/N_f$ fixed.  In this limit, $a_{SCFT}(N_c, N_f)\to N_f^2 \widehat a _{SCFT}(x).$  

We now consider the Wilsonian flow associated with $\ev{M}$ for $M=Q\tilde Q$, 
\eqn\flowe{\ev{Q_{N_f, N_c}}=\ev{\tilde Q_{\tilde N_f}^{N_c}}=v, \qquad \ev{X_i}=0: \qquad (N_c, N_f)\to (N_c-1, N_f-1+N_a),}
where one flavor is eaten and $N_a$ additional flavors come from decomposing under $SU(N_c)\to SU(N_c-1)$; $X\to \widehat X+Q_X+\tilde Q_X +S$, with $S$ a singlet.  The Higgsed theory has, all together, $2N_f-1+N_a$ singlets, which are IR-free fields (at point (A) in Fig. 2).  The a-theorem for this flow thus states (including $a_{free}={2\over 9}$ for each free singlet)
\eqn\aineq{a_{SCFT}(N_c, N_f)\geq a_{SCFT}(N_c-1, N_f-1+N_a)+{2\over 9}(2N_f-1+N_a).}
Taking the Veneziano limit, with $x\equiv N_c/N_f$ fixed and $\widehat a\equiv a/N_f^2$, this gives
\eqn\aineqlim{2(1-N_a)\widehat a (x)+((N_a-1)x +1){d\over dx}a(x)\geq {4\over 9}.}
The inequality is satisfied, $\Delta a<0$, with $\Delta a\neq 0$, so $\ev{M}$ is relevant.  The fields here have $R_{micro}(Q_f)>0$ and $R_{micro}(X_i)>0$, so these examples  fit with the frequent correlation between relevant Higgsing and $R_{micro}>0$.  

\newsec{Examples of irrelevant / marginal $\vev{M}$, $\Delta a=0$,  with an interacting sector}

We now consider the $A_k$ theories \refs{\Kutasov, \Kutasovii, \Kutasoviii}: $SU(N_c)$ SQCD with $N_f$ flavors, adjoint $X$, and  $W_{elec}=\Tr X^{k+1}$, which has a $SU(\tilde N_c\equiv kN_f-N_c)$ magnetic dual with 
\eqn\kutdual{W_{dual}=-\Tr Y^{k+1} +\sum _{j=1}^k M_j \tilde q Y^{k-j}q,}
where $M_{j=1\dots k}$ map to the generalized mesons $M_j\to \tilde Q X^{j-1}Q$ on the electric side.  A-maximization is used  \KPS\ to determine when the $A_k$ theories exist, by  analyzing when $W_{elec}$ is a relevant deformation of the $\widehat A$ SCFTs.    We recall some results, referring the reader to \KPS\ for more details.  Again, it simplifies the expressions to take the Veneziano limit of large $N_f$ and $N_c$, holding fixed  $x\equiv N_c/N_f$.  The electric theory is asymptotically free for $x>x_{FE}=\half$, and its vacuum stability (avoiding $W_{dyn}\neq 0$) requires $x<k$.  The electric description is weak for $x\approx \half$, and becomes more strongly coupled for larger $x$.  The magnetic description is weak for $\tilde x\equiv k-x\approx \half$, and becomes more strongly coupled for larger $\tilde x$.  In cases where the two descriptions seem to disagree, given that we trust the duality, the more weakly coupled description is more reliable and presumably correct.

The $A_k$ SCFT only exists for $x>x_k$, i.e. where $R(X^{k+1})\leq 2$ in the $\widehat A$ SCFT, to drive relevant $\widehat A\to A_k$ RG flow \KPS.  For the range $x_k<x<k$, the $A_k$ theory has   
\eqn\akr{A_k:\ R_{micro}(X)={2\over k+1}, \qquad\hbox{and}\qquad R_{micro}(Q)={k+1-2x \over k+1}.}So $R_{micro}(Q)<0$ for $x>(k+1)/2$, which will be a case of interest  here.   Using \akr, 
\eqn\rmicro{R_{micro}(M_j)=R_{micro}(\tilde Q X^{j-1}Q)={2(k+j-2x)\over k+1}.}
When $R_{micro}(M_j)<{2\over 3}$, the unitarity bound requires that $M_j$ is IR free.  

There can be additional accidental symmetries, seen only from the duality.  Depending on $x$, various term in the dual \kutdual\ are relevant or irrelevant.  An extreme case is $x\geq x_{FM}\equiv k-\half$, where the entire dual $SU(\tilde N_c)$ theory is IR free.  For $x$ just below $x_{FM}$, the theory is in a magnetic Banks-Zaks limit, where the $SU(\tilde N_c)$ is barely interacting and $R\approx 2/3$ for every field, so every non-cubic term in \kutdual\ is irrelevant;  the meson $M_k$ in \kutdual\ has the weak interaction with $q\tilde q$ in \kutdual, while all mesons $M_{j<k}$ are IR free. For the window $x_k<x<k-\tilde x_k$ \KPS, where $\Tr X^{k+1}$ and $\Tr Y^{k+1}$ are both relevant deformations of the electric and magnetic $\widehat A$ SCFTs, $R_{micro}(M_j)+R(\tilde q Y^{k-j}q)=2$ and the $M_j$ term in \kutdual\ is irrelevant precisely when $R_{micro}(M_j)\leq {2\over 3}$.   More generally, a-maximization is required on both sides \KPS.

We now consider taking $\vev{M_j}=\vev{\tilde Q X^{j-1}Q}\neq 0$.  The case of $M_{j=k}=\tilde QX^{k-1}Q$ expectation value, which maps to a $\tilde q q$ mass term in the dual \kutdual, was considered as a check of the duality in the original works \refs{\Kutasov, \Kutasovii, \Kutasoviii}.  It is clear from the dual that $\vev{M_k}$ leads to $\Delta a\neq 0$, since it sources a $\tilde q q$ mass term, which is always relevant.     On the other hand, as seen from the dual \kutdual,  the other $\vev{M_{j<k}}$ can source irrelevant interactions.

If $\ev{M_j}$ sources an irrelevant interaction, the magnetic description reveals physics that is very different than what would have been expected from the electric description, or from classical intuition about Higgsing removing d.o.f..  The magnetic description shows that in such cases  $\ev{M_j}$ is a freely generated subspace of the moduli space, with $\Delta a=0$ everywhere on this space.   A necessary condition for this to happen is that $\ev{M_j}$ does not fully break the gauge group, both to avoid classical constraints on $\ev{M_j}$, and also to avoid having $\ev{M_j}$ connect to where the quantum effects can be made small, i.e. fully Higgsing the gauge group with large vevs.  In cases where $\Delta a=0$, evidently never-small quantum effects on the $\ev{M_j}$ moduli spaces must eliminate the non-zero $\Delta a_{classical}$.

Let us consider the extreme case of $\vev{M_{j=1}}\neq 0$ in more detail, first in the electric description.  As in \flowe, we take $\ev{X}=0$ and  $\ev{Q_{N_f, N_c}}=\ev{\tilde Q_{\tilde N_f}^{N_c}}=v$, which breaks $SU(N_c)\to SU(N_c-1)$, with $N_f\to N_f-1+1$, with the $+1$ 
additional flavor from decomposing the $SU(N_c)$ adjoint under $SU(N_c-1)$, $X \to \widehat X+Q_X+\tilde Q_X+S$, with $S$ a singlet, with interactions 
(not bothering with numerical factors) 
\eqn\wpert{W_{tree}=\Tr X^{k+1}\to \Tr \widehat X^{k+1}+ \widetilde Q_X \widehat X^{k-1}Q_X+S\Tr \widehat X^k+S^{k+1}+\dots.}  
For $x<x_k$ all terms in \wpert\ are irrelevant, so we can ignore $W$ before and after Higgsing,
\eqn\aiaf{a_i=a_{\widehat A}(N_c, N_f), \qquad a_f=a_{\widehat A}(N_c-1, N_f)+{2\over 9}(2N_f), \qquad \hbox{for} \qquad x<x_k,}
as in \aineq\ for $N_a=1$.   Here $a_i$ is computed in the $SU(N_c)$ $\widehat A$ SCFT with $v=0$, using a-maximization as in  \KPS, while  $a_f$ is computed in the Higgsed, $v\neq 0$, $SU(N_c-1)$ IR theory.   For 
$x>x_k$ the $W_{A_k}=\Tr X^{k+1}$ term in \wpert\ is relevant for $a_i$, and the $W_{A_k, M_k}=\Tr \widehat X^{k+1} + (M_k)_{N_fN_f}$, with $ (M_k)_{N_f,N_f}=\tilde Q_X X^{k-1}Q_X$, terms are relevant for $a_f$, with the other terms irrelevant.  The effect of the mesonic $\Delta W= (M_k)_{N_f,N_f}$ is a RG flow from $A_k$ to a new IR fixed point, where $R(Q_X)=R(X)=2/(k+1)$, so 
\eqn\aiaff{a_i=a_{A_k}(N_c, N_f), \qquad a_f=a_{A_k, M_k}(N_c-1, N_f-1,1)+{2\over 9}(2N_f), \qquad \hbox{for} \qquad x>x_k.}

We can now compute $\Delta a\equiv a_f-a_i$ for the Higgsing, using \aiaf\ or \aiaff.  The result is found to satisfy $\Delta a\leq 0$, with $\Delta a=0$ occurring at $x=(k+1)/2$, which is precisely where $R(Q)=0$.  Since $x_k<(k+1)/2$ \KPS, this is in the region of the $A_k$ SCFT, so \aiaff\ applies. Indeed, the $\widehat A$ SCFT has $R(Q)>0$, and \aiaf\ yields $\Delta a<0$.

On the magnetic side, things are much simpler, since the $\ev{M_1}$ deformation triggers
\eqn\magdef{\Delta W_{dual}= \ev{M_1} \tilde q Y^{k-1}q:  \qquad\hbox{irrelevant if}\quad  R(\tilde q Y^{k-1}q)>2,}
and relevant otherwise.  The relevant case certainly leads to $\Delta a<0$, in accord with the a-theorem.   The irrelevant case, on the other hand,  leads to $\Delta a\to 0$ in the IR. 

The effect of $\vev{M_1}$ for $k=2$ was discussed in some detail in \AharonyNE, where the effect of $\vev{M_1}\tilde q Y q$ was analyzed by considering a UV completion, to replace the quartic term in $W_{dual}$ with purely cubic terms, with some additional massive flavors integrated-in, such that integrating them back out gives back the quartic terms.  Likewise, one could here UV complete $\Delta W_{IR}\supset M_j \widetilde q Y^{k-j}q$ to a theory with only cubic interactions, by integrating-in  $k-j$ additional pairs of massive matter fields $F_i$, $\tilde F_i$ and $G_j$, $\tilde G_j$, with $W_{UV}\supset M_j q \tilde F_1+F_1Y\tilde F_2+ F_2 Y\tilde F_3+m\sum _i F_i\tilde F_i$.  Then $M_j\to \vev{M_j}$ leads to mass mixing, which can be re-diagonalized, similar to \AharonyNE.   But for our interests here, such UV completions are not needed, and the IR effect can just as well be obtained directly from \magdef.

To determine $R(\tilde q Y^{k-j}q)$, we must account for the relevance or irrelevance of the various terms in  \kutdual, including $\Tr Y^{k+1}$ \KPS.  The a-maximization procedure is best implemented by computer, but we can make some general, qualitative remarks.   For $x\approx x_{FM}=k-\half$ the theory is in 
the weakly coupled magnetic Banks-Zaks regime, $\Tr Y^{k+1}$ is irrelevant (for $k>2$), and $R(q)=R(\tilde q)\approx R(X)\approx {2\over 3}^-$, so $M_k$ is weakly coupled and  all $M_{j<k}$ are IR free; then deforming by $\ev{M_k}$ is relevant, deforming by $\ev{M_{k-1}}$ is weakly relevant, and any other $\ev{M_{j<k-1}}$ source irrelevant interactions, leading to $\Delta a=0$.   The extreme case is $\ev{M_1}$, which sources $\tilde q  Y ^{k-1}q$, which for $k>2$ is certainly irrelevant for sufficiently small $\tilde x$.  As we increase $\tilde x$, it can be seen more generally that $R(\tilde q Y^{k-1}q)>R(Y^{k+1})$, since $R(q)>R(Y)$, so $\Tr Y^{k+1}$ becomes relevant before $\ev{M_1}$. 

This is nice: the critical $\tilde x$, where $\ev{M_1}$ crosses from sourcing an irrelevant to relevant $\Delta W$ in the magnetic description, is where $\Tr Y^{k+1}$ is already relevant, so a-maximization is not needed there.  We can simply use $R(Y)=2/(k+1)$, which determines $R(\tilde q Y^{k-1}q)=2-R_{micro}(M_1)$.  Therefore,  using \magdef, $\ev{M_1}$ becomes irrelevant on the magnetic side precisely when $R_{micro}(M_1)<0$, i.e. when the electric theory has $R_{micro}(Q)\leq 0$.    

So the electric and the magnetic dual descriptions both give $\Delta a=0$ at $x=(k+1)/2$, where $R_{micro}(Q)=0$, but they seemingly disagree for $x>(k+1)/2$.  The detailed analysis on the electric side (accounting for all visible accidental symmetries)  gives $\Delta a\neq 0$ for $x>(k+1)/2$.  But the magnetic analysis gives $\Delta a=0$ for all $x\geq (k+1)/2$, since $\ev{M_1}$ sources $\tilde q Y^{k-j}q$, which is there irrelevant.   The magnetic result is presumably correct (assuming the duality is correct), since it is more weakly coupled at large $x$. Evidently, the magnetic dual reveals accidental symmetries that were not evident in the electric description, giving $\Delta a=0$ for $x\geq (k+1)/2$, rather than just at $x=(k+1)/2$.

There are many analogous examples, e.g. using the $D_k$ and $E_k$ theories of \IntriligatorMI, which we have checked are quite similar to the $A_k$ case.

\centerline{\bf Acknowledgments}

We would like to thank Jeff Fortin for discussions and initial participation in this project.   KI thanks Ofer Aharony, Zohar Komargodski, David Kutasov, and Adam Schwimmer for discussions, and the Weizmann Institute Theory Group, and the IHES, for hospitality while some of this work was done.  He also thanks the organizers of workshops at PI, McGill, and ENS Paris, for the chance to talk about this, and the participants for their comments. This work was supported by DOE-FG03-97ER40546.

\listrefs

\end